\documentclass[usenatbib]{mn2e}

\usepackage{times}
\usepackage{natbib}
\usepackage{graphicx}
\usepackage{amsmath} 
\usepackage{amssymb}
\title[]{Diffuse radio emission in MACS~J1752.0+4440}
\author[R.~J. van Weeren et al.]{R.~J. van Weeren$^{1,2}$\thanks{E-mail:
rvweeren@strw.leidenuniv.nl},  A.~Bonafede$^{3}$, H. Ebeling$^{4}$, A.~C. Edge$^{5}$, M. Br\"uggen$^{3,6}$, \newauthor
 G. Giovannini$^{7,8}$,   M. Hoeft$^{9}$,  and H.~J.~A. R\"ottgering$^{1}$ \\ \\ 
$^{1}$Leiden Observatory, Leiden University, P.O. Box 9513, NL-2300 RA Leiden, The Netherlands\\
$^{2}$Netherlands Institute for Radio Astronomy (ASTRON), Postbus 2, 7990 AA Dwingeloo, The Netherlands\\
$^{3}$Jacobs University Bremen, P.O. Box 750561, 28725 Bremen, Germany \\
$^{4}$ Institute for Astronomy, University of Hawaii, 2680 Woodlawn Drive, Honolulu, HI 96822, USA  \\
$^{5}$ Institute for Computational Cosmology, Department of Physics, Durham University, Durham, DH1 3LE, UK \\
$^{6}$ Hamburger Sternwarte, Gojenbergsweg 112, 21029 Hamburg, Germany \\
$^{7}$ Dipartimento di Astronomia, via Ranzani 1, 40127 Bologna, Italy \\
$^{8}$ Istituto di Radioastronomia-INAF, via P.Gobetti 101, 40129 Bologna, Italy \\
$^{9}$Th\"uringer Landessternwarte Tautenburg, Sternwarte 5, 07778, Tautenburg, Germany \\}

\begin{document}

\date{ } 

\pagerange{\pageref{firstpage}--\pageref{lastpage}} \pubyear{2011}

\maketitle

\def\aj{{AJ}}                   
\def\araa{{ARA\&A}}             
\def\apj{{ApJ}}                 
\def\apjl{{ApJ}}                
\def\apjs{{ApJS}}               
\def\ao{{Appl.~Opt.}}           
\def\apss{{Ap\&SS}}             
\def\aap{{A\&A}}                
\def\aapr{{A\&A~Rev.}}          
\def\aaps{{A\&AS}}              
\def\azh{{AZh}}                 
\def\baas{{BAAS}}               
\def\jrasc{{JRASC}}             
\def\memras{\ref@jnl{MmRAS}}            
\def\mnras{{MNRAS}}             
\def\pra{\ref@jnl{Phys.~Rev.~A}}        
\def\prb{\ref@jnl{Phys.~Rev.~B}}        
\def\prc{\ref@jnl{Phys.~Rev.~C}}        
\def\prd{\ref@jnl{Phys.~Rev.~D}}        
\def\pre{\ref@jnl{Phys.~Rev.~E}}        
\def\prl{\ref@jnl{Phys.~Rev.~Lett.}}    
\def\pasp{{PASP}}               
\def\pasj{{PASJ}}               
\def\qjras{\ref@jnl{QJRAS}}             
\def\skytel{\ref@jnl{S\&T}}             
\def\solphys{\ref@jnl{Sol.~Phys.}}      
\def\sovast{\ref@jnl{Soviet~Ast.}}      
\def\ssr{\ref@jnl{Space~Sci.~Rev.}}     
\def\zap{\ref@jnl{ZAp}}                 
\def\nat{{Nature}}              
\def\iaucirc{\ref@jnl{IAU~Circ.}}
\def\aplett{\ref@jnl{Astrophys.~Lett.}}
\def\apspr{\ref@jnl{Astrophys.~Space~Phys.~Res.}}
\def\bain{\ref@jnl{Bull.~Astron.~Inst.~Netherlands}}
\def\fcp{\ref@jnl{Fund.~Cosmic~Phys.}}
\def\gca{\ref@jnl{Geochim.~Cosmochim.~Acta}}
\def\grl{\ref@jnl{Geophys.~Res.~Lett.}}
\def\jcp{\ref@jnl{J.~Chem.~Phys.}}      
\def\jgr{\ref@jnl{J.~Geophys.~Res.}}    
\def\jqsrt{\ref@jnl{J.~Quant.~Spec.~Radiat.~Transf.}}
\def\memsai{\ref@jnl{Mem.~Soc.~Astron.~Italiana}}
\def\nphysa{\ref@jnl{Nucl.~Phys.~A}}
\def\physrep{{Phys.~Rep.}}
\def\physscr{\ref@jnl{Phys.~Scr}}
\def\planss{\ref@jnl{Planet.~Space~Sci.}}
\def\procspie{\ref@jnl{Proc.~SPIE}}
\let\astap=\aap
\let\apjlett=\apjl
\let\apjsupp=\apjs
\let\applopt=\ao

\label{firstpage}

\begin{abstract}
We report the discovery of large-scale diffuse radio emission in the galaxy cluster MACS~J1752.0+4440  ($z=0.366$). 
Using Westerbork Synthesis Radio Telescope (WSRT) observations we find that the cluster hosts a double radio relic system as well as a 1.65~Mpc radio halo covering the region between the two relics. The relics are diametrically located on opposite sides of the cluster center. The NE and SW relics have sizes of 1.3 and 0.9~Mpc, respectively.  In case of an isolated binary merger event, the relative sizes of the relics  suggest a mass ratio about $2 : 1$. We measure integrated spectra of $-1.16\pm0.03$ for the NE and $-1.10\pm0.05$ for the SW relic. We conclude that this cluster has  undergone a violent binary merger event and the relics are best explained by particles (re)accelerated in outwards traveling shock waves. The spectral indices suggest the relics trace shock waves with Mach numbers  ($\cal{M}$) of around 3.5 to 4.5. These relatively high Mach numbers derived from the radio spectral index are comparable to those derived for a few other recently discovered relics. This implies that merger shocks with $\mathcal{M} > 3$ are relatively common in cluster outskirts if our understanding of diffusive shock acceleration is correct.

\end{abstract}


\section{Introduction}
Giant radio relics are extended radio sources found in the outskirts of disturbed galaxy clusters. Particularly interesting are so-called double radio relics which are found on opposites sides of the cluster center and are believed to trace two shock waves created in the intracluster medium (ICM) during a binary merger event. In the case of advantageous inclination with respect to our line of sight the location of the relics makes it possible to directly constrain the merger geometry and together with modeling can be used to derive the mass ratio and impact parameter of the merger event \citep{1999ApJ...518..603R, 2011MNRAS.418..230V}.  

\begin{table*}
\begin{center}
\caption{WSRT observations}
\begin{tabular}{llllll}
\hline
\hline
& 25 cm & 21 cm & 18 cm  & 13 cm \\
\hline
Bandwidth, \# channels per subband            & $8\times20$~MHz, 64 & $8\times20$~MHz, 64  & $8\times20$~MHz, 64  & $8\times20$~MHz, 64 & \\
Observation dates	      &  November 10, 2011 &7 \& 12  March, 2011& May 23 \& June 16, 2011 & July 5 \& 6, 2011 &\\ 
Total on-source time     & 10~hr&12~hr&24~hr&24~hr\\
Beam size                      & $28.9\arcsec\times16.6\arcsec$ & $27.1\arcsec\times17.0\arcsec$  &$17.3\arcsec \times 11.8\arcsec$,  $23.5\arcsec \times 15.9\arcsec$ &$12.7\arcsec\times8.9\arcsec$  \\
Rms noise ($\sigma_{\rm{rms}}$) &35~$\mu$Jy~beam$^{-1}$ & 26~$\mu$Jy~beam$^{-1}$ &  16, 18~$\mu$Jy~beam$^{-1}$ & 23~$\mu$Jy~beam$^{-1}$& \\
Robust weighting &1.0 & 1.0&0.0, 3.0 &0.0\\
\hline
\hline
\end{tabular}
\label{tab:wsrtobservations}
\end{center}
\end{table*}

Shock waves in galaxy clusters are proposed sites for particle acceleration, for example via the diffusive shock acceleration (DSA)  mechanism \citep{1983RPPh...46..973D, 1987PhR...154....1B, 1991SSRv...58..259J}. However, the efficiency with which collisionless shocks can accelerate particles might not be high enough  to produce the observed radio brightness \citep[e.g.,][]{2011ApJ...728...82M}. The diffusive shock acceleration may be aided by the presence of relativistic particles in the upstream plasma \citep[e.g.,][]{2005ApJ...627..733M, 2008A&A...486..347G, 2011ApJ...734...18K}.  Relics are expected to be preferably found in  cluster outskirts because the kinetic energy dissipated in merger shocks peaks at a distance of about 1~Mpc from the cluster center \citep{2012MNRAS.421.1868V}. 

Radio halos are another class of diffuse extended radio sources found in merging clusters \citep[e.g.,][]{2010ApJ...721L..82C}. They have Mpc-sizes and are centrally located.  Halos  have been explained by in-situ acceleration of particles through merger generated turbulence \citep[][]{2001MNRAS.320..365B, 2001ApJ...557..560P} or by secondary electrons which are continuously injected into the ICM by inelastic collisions between relativistic and thermal protons \citep[][]{1980ApJ...239L..93D, 1999APh....12..169B, 2000A&A...362..151D}.  Secondary models are currently disfavored as they do not explain the existence of ultra-steep spectrum radio halos \citep{2008Natur.455..944B}, and so far no Gamma ray emission has been detected by the Fermi satellite at the levels expected \citep{ 2011ApJ...728...53J}. Clusters also show a bimodal distribution in the $P_{\rm{1.4~GHz}} - L_{\rm{X}}$ diagram, with only a minority of clusters hosting halos \citep{2007ApJ...670L...5B}, in agreement with the turbulent re-acceleration model \citep{ 2009A&A...507..661B}. However, \cite{2011A&A...527A..99E} claim  that cosmic ray (CR) transport in clusters can also induce the radio bimodality as merging clusters should have a much more centrally concentrated CR population than relaxed ones, resulting in higher  synchrotron emissivities. Basic turbulent re-acceleration models predict steeping of the radio spectrum at higher-frequencies \citep[e.g., fig.~1 in][]{2010A&A...509A..68C}. Recent LOFAR observations revealed that the radio spectrum of the Abell 2256 halo steepens at low frequencies. This unexpected result shows that the formation process of halos might be more complex than previously thought \citep[][]{2012arXiv1205.4730V}.
 
Although more than 30 relics have been found, they are mostly located in clusters at $z \lesssim 0.3$, while based on simulations more relics at larger redshift are expected \citep{2012MNRAS.420.2006N}. 
According to \citeauthor{2012MNRAS.420.2006N}, the fraction of relics in X-ray selected cluster samples as function of redshift gives a powerful tool to constrain the evolution of magnetic fields and  particle acceleration in the ICM . 
 The few known $z>0.3$ relics are MACS~J0717.5+3745 \citep{2003MNRAS.339..913E, 2009A&A...503..707B, 2009A&A...505..991V}, PLCK G287.0+32.9 \citep{2011ApJ...736L...8B, 2011A&A...536A...8P} and CL1446+26 \citep{2009A&A...507.1257G}.  The extreme merger event of MACS~J0717.5+3745 has been studied extensively \citep{2003ApJ...583..559L, 2004ApJ...609L..49E, 2008MNRAS.389.1240K, 2008ApJ...684..160M, 2009ApJ...693L..56M, 2009ApJ...707L.102Z, 2011MNRAS.410.2593M,2011arXiv1109.3301L}, but the systems complexity also makes the interpretation of the results more difficult. 

We embarked on a search for less complex, but still  massive merging clusters in the MACS sample of X-ray luminous clusters at $z > 0.3$ \citep{2001ApJ...553..668E, 2012MNRAS.420.2120M}.  A candidate for a $z> 0.3$ double relic system was reported by \cite{2003MNRAS.339..913E} in MACS~J1752.0+4440 with two radio sources found on opposite sides of the cluster center in NRAO VLA Sky Survey \citep[NVSS,][]{1998AJ....115.1693C} and Westerbork Northern Sky Survey \citep[WENSS,][]{1997A&AS..124..259R} images. Here we present WSRT and archival Very Large Array (VLA) observations of this cluster. In Sect.~\ref{sec:observations} we report on the radio observations and data reduction. The results are presented in Sect.~\ref{sec:results}. We end with a discussion and summary in Sects.~\ref{sec:discussion} and \ref{sec:summary}. Throughout this paper we assume a $\Lambda$CDM cosmology with $H_{0} = 71$~km~s$^{-1}$~Mpc$^{-1}$, $\Omega_{m} = 0.27$, and $\Omega_{\Lambda} = 0.73$. At the redshift of MACS~J1752.0+4440 ($z=0.366$) $1\arcsec$~corresponds to 5 kpc. All images are in the J2000 coordinate system.

\section{Observations}
\label{sec:observations}

MACS~J1752.0+4440 was observed with the WSRT in the L-band at 25, 21, 18~cm and in the S-band at 13~cm, see Table~\ref{tab:wsrtobservations} for an overview. The data were calibrated with CASA. As a first step we removed time ranges affected by shadowing. Radio frequency interference (RFI) was flagged with the AOFlagger from \cite{2010MNRAS.405..155O}. The data were corrected for the bandpass response and gain solutions were determined for the calibrator sources and transferred to the target source. The flux densities for the calibrator sources were set by the \cite{perleyandtaylor} extension to the \cite{1977A&A....61...99B} scale.  The data were further self-calibrated in the NRAO Astronomical Image Processing System (AIPS) package.  Images were cleaned with manually placed clean boxes and corrected for the primary beam attenuation.

Archival VLA L-band B-array observations  (project AE147; July 7, 2002) were obtained from the NRAO VLA Archive Survey (NVAS). From these data we created an image with a resolution of $5.1\arcsec~\times~4.5\arcsec$ (robust 1.0 weighting) and a noise of 28~$\mu$Jy~beam$^{-1}$. The resolution and noise levels for the WSRT images are reported in Table~\ref{tab:wsrtobservations}.

\begin{figure*}
\includegraphics[angle =90, trim =0cm 0cm 0cm 0cm,width=0.49\textwidth, clip=true]{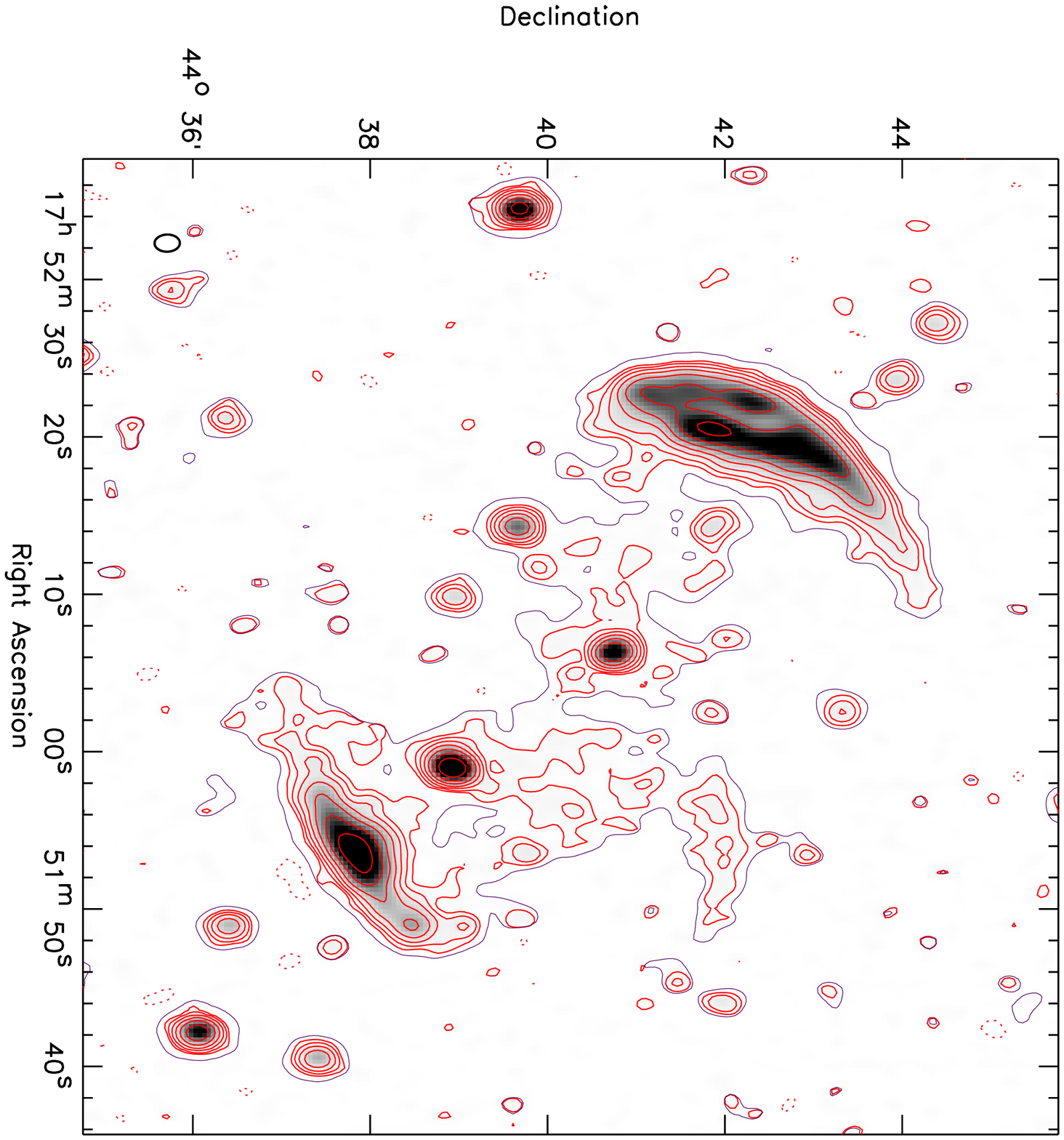}
\includegraphics[angle =90, trim =0cm 0cm 0cm 0cm,width=0.49\textwidth, clip=true]{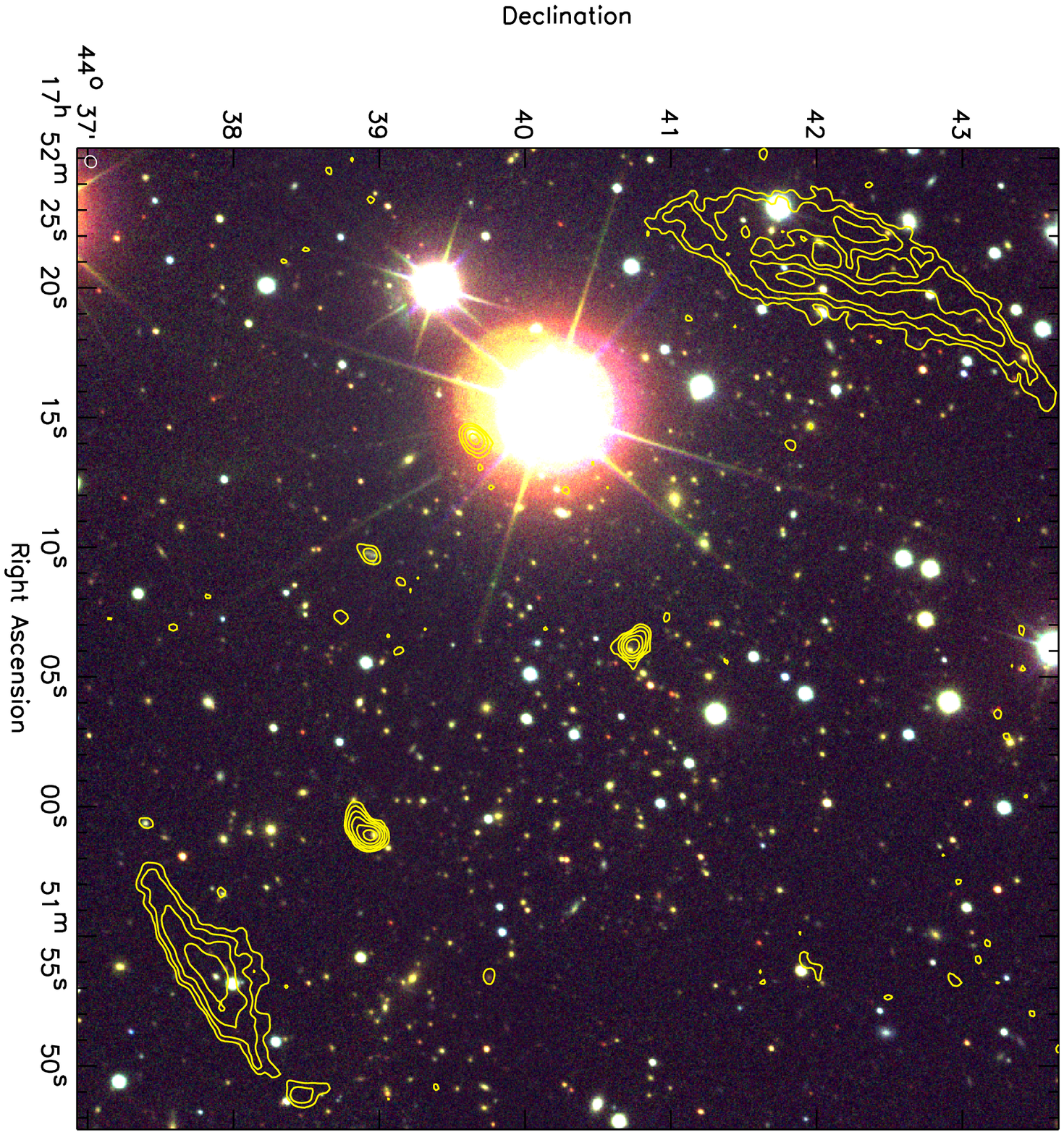} 
\caption{Left: WSRT 18~cm image. High-resolution image made with robust~=~0 weighting. Red contour levels are drawn at $[1, 2, 4, \ldots] \times 3\sigma_{\mathrm{rms}}$.  Negative  $-3\sigma_{\mathrm{rms}}$ contours are shown by the dotted lines. Blue contours drawn at 54~$\mu$Jy~beam$^{-1}$ are from a lower resolution   image made with robust~=~3 weighting to enhance diffuse emission. Right: SDSS DR7 color image (\emph{g}, \emph{r}, and \emph{i} bands) of the cluster center overlaid with VLA radio contours. 
Contour levels are drawn at  $[1, 2, 4, \ldots] \times 3\sigma_{\mathrm{rms}}$. 
} 
\label{fig:18cm}
\end{figure*}

\section{Results}
\label{sec:results}
\begin{figure*}
\includegraphics[angle =90, trim =0cm 0cm 0cm 0cm,width=0.32\textwidth, clip=true]{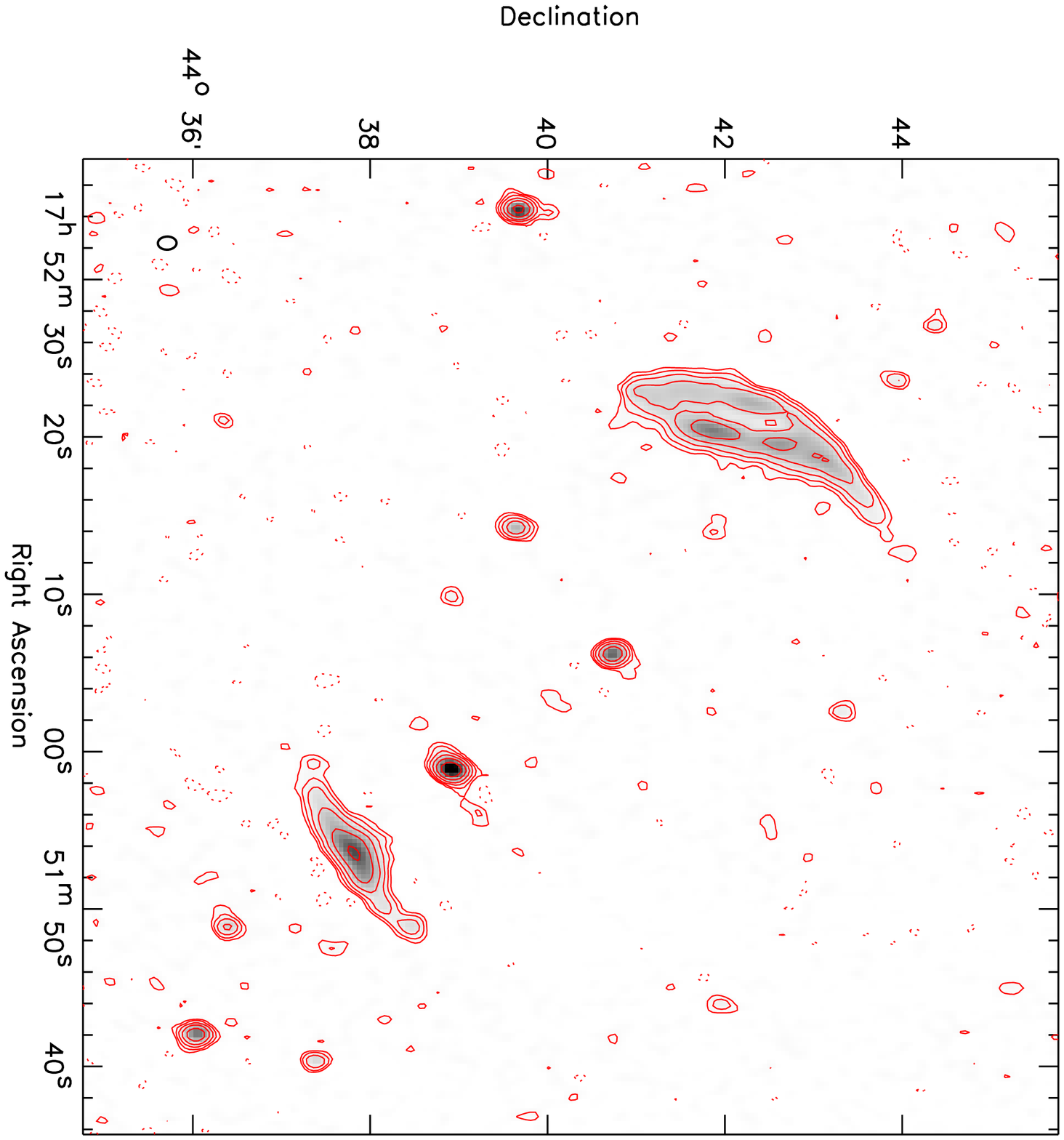}
\includegraphics[angle =90, trim =0cm 0cm 0cm 0cm,width=0.32\textwidth, clip=true]{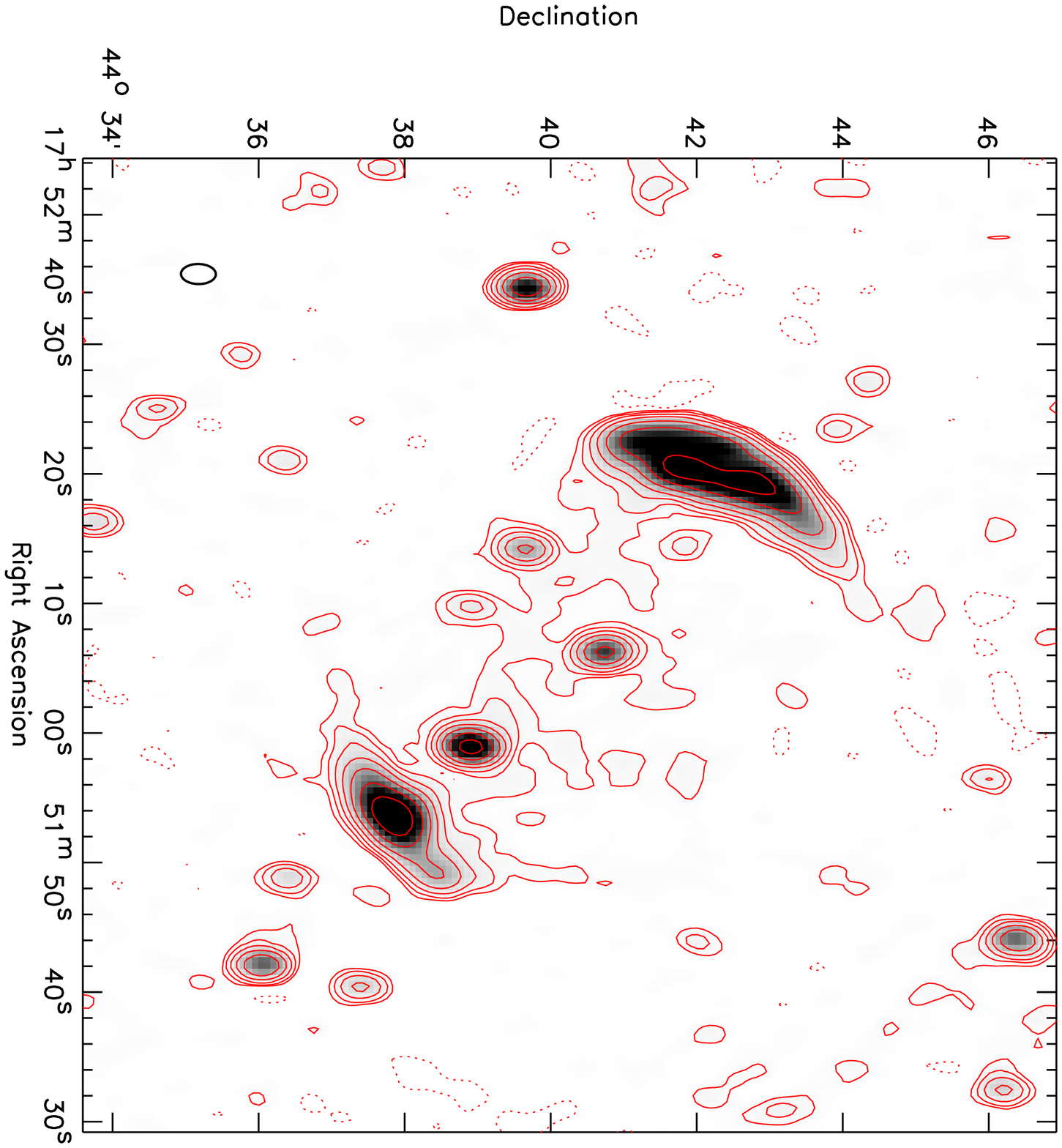}
\includegraphics[angle =90, trim =0cm 0cm 0cm 0cm,width=0.32\textwidth, clip=true]{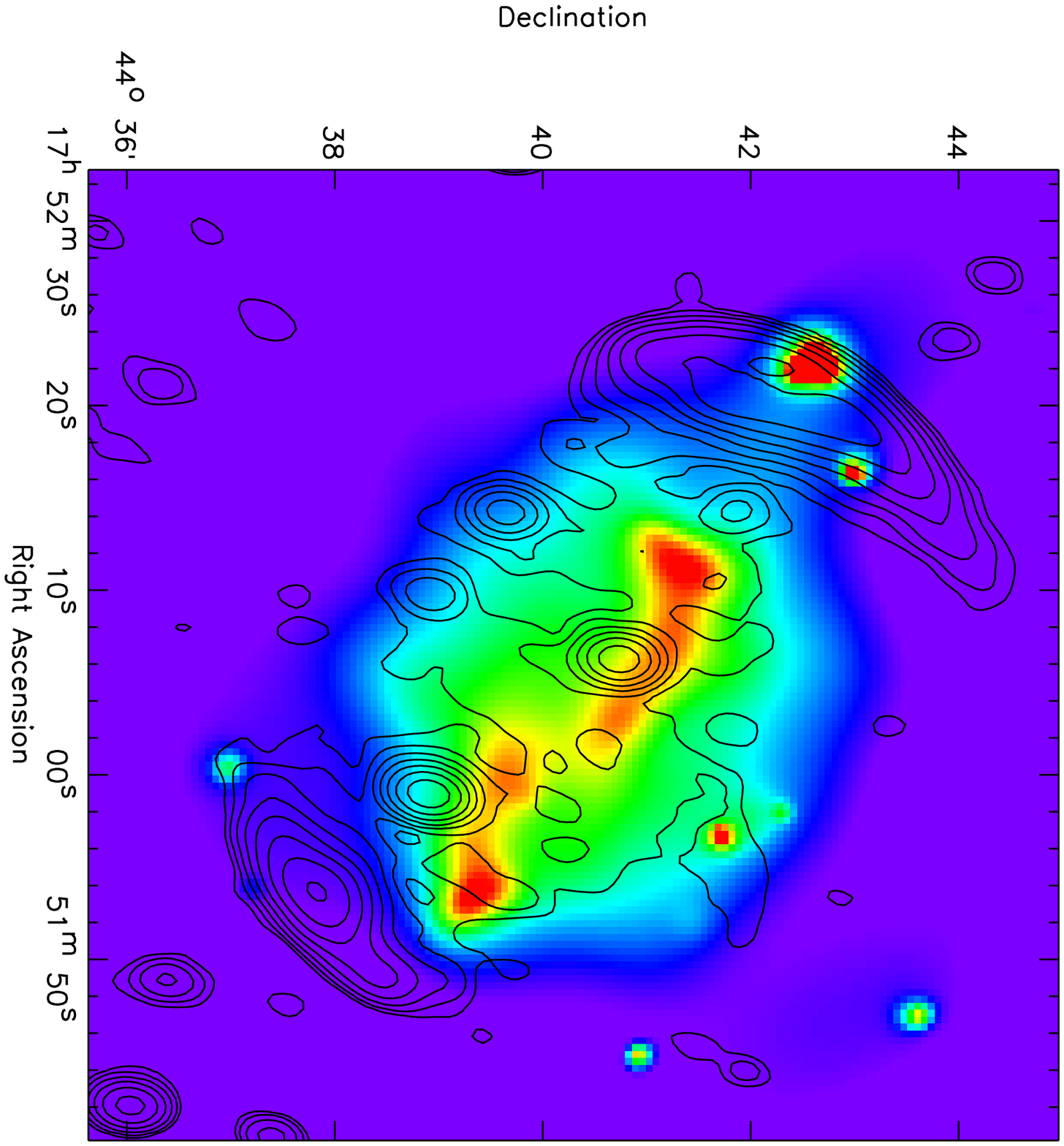}
\caption{WSRT 13 (left) and 25 (middle) cm images. The beam sizes are  shown in the bottom left corner of the images, see also Table~\ref{tab:wsrtobservations}. Contour levels are drawn as in Fig.~\ref{fig:18cm}. Right: Adaptively smoothed XMM-Newton MOS 0.2--12.0~keV image (\citeauthor{ebeling}). Contours are from the WSRT low-resolution 21~cm image and drawn as in Fig. ~\ref{fig:18cm}.} 
\label{fig:vla}
\end{figure*}

MACSJ1752.0+4440 was identified as a cluster at $z=0.366$ by  \cite{2003MNRAS.339..913E} based on its detection in the ROSAT All-Sky Survey  \citep[RASS,][]{1999A&A...349..389V}. At $L_{\rm X,~0.1-2.4~keV} = 8.2\times 10^{44}$ erg s$^{-1}$, MACS~J1752.0+4440 is one of the most X-ray luminous clusters known at $z>0.3$. And SDSS image is shown in Fig.~\ref{fig:18cm}  \citep{2009ApJS..182..543A}. The WSRT 18~cm image reveals two arc-like radio sources on opposite sides of the cluster center (Fig.~\ref{fig:18cm}). Additional faint diffuse emission is present between these two sources. In the 21 and 25~cm WSRT images we also detect this low surface brightness emission, although with a lower signal to noise ratio (SNR), see Fig.~\ref{fig:vla}. The brighter arc-like sources are also detected in VLA 1.4~GHz B-array and WSRT 13~cm images (Figs.~\ref{fig:18cm} and \ref{fig:vla}). An XMM-Newton image (Fig.~\ref{fig:vla}) reveals a disturbed and elongated ICM. \citeauthor{ebeling} will present an analysis of these X-ray observations. 

There is a faint optical object (probably a galaxy) visible in the SDSS image about 5\arcsec to the south of a foreground star, at the center of the SW arc-like radio source. However, it is unlikely that the SW radio source can be classified as a radio galaxy because there is no compact radio core visible, which is expected for an active AGN.
% RELICS 
We classify the two arc-like sources as a double radio relic system because of (i) their morphology and sizes, (ii) location with respect to the cluster center,  (iii) location along the merger axis, and (iv) perpendicular orientation with respect to the merger axis.

The NE and SW relics feature extents of 1.3 and 0.9 Mpc, respectively (Table~\ref{tab:macs1752properties}). The relics have well defined sharp outer edges, while their surface brightness drops somewhat more gradually towards the cluster center, as is seen for most double radio relics. The NE relic is composed of two linear filaments.  Across the length of the NE relic the emission fades more slowly towards the north than at the relics's southern end.

% HALO
The diffuse emission between the radio relics is classified as a 1.65~Mpc extended radio halo because the radio emission follows the X-ray emission from the ICM. Interestingly, a faint EW elongated smudge of emission is visible at the northern end of the radio halo, see Fig.~\ref{fig:18cm}. It could be a fainter radio relic but this is somewhat speculative as the feature has a low-surface brightness. 

The flux densities for the two relics and radio halo are reported in Table~\ref{tab:macs1752properties}. The two brightest compact sources between the two relics are probably AGN in the cluster, based on the colors of their optical counterparts visible in the SDSS image. They are slightly extended ($\sim10-15\arcsec$) in the VLA-B array image. We obtained the integrated flux density of the halo from the 18~cm data as the halo is most clearly visible in this band (due to the longer integration time). We first subtracted the flux contribution from compact sources by creating a high-resolution image using uniform weighting and an inner uv-range cut of 200~m. The found clean components were subtracted from the uv-data and then we re-imaged the data with natural weighting. The halo flux density was calculated by summing over the entire area between the two relics. This have a flux density of $11.4\pm1.1$~mJy, corresponding to a 1.4~GHz radio halo luminosity $P_{1.4 \rm { GHz}}=9.5 \times 10^{24}$~W~Hz$^{-1}$ (assuming a spectral index of $-1.1$). The halo linear size and radio power are in agreement with the distribution presented in \cite{2009A&A...507.1257G}. The X-ray luminosity and radio power are consistent with the $L_{\rm{X}} - P_{1.4 \rm { GHz}}$ correlation for radio halos \citep[e.g.,][]{2000ApJ...544..686L, 2002A&A...396...83E, 2006MNRAS.369.1577C,2007MNRAS.378.1565C,2009A&A...507..661B,2009A&A...499..679M,2012arXiv1205.1919F}. Although the halo is also detected in the 21 and 25~cm images, the SNR is too low and frequency span too small to obtain a useful spectral index estimate.

\begin{figure}
\includegraphics[angle =90, trim =0cm 0cm 0cm 0cm,width=0.49\textwidth, clip=true]{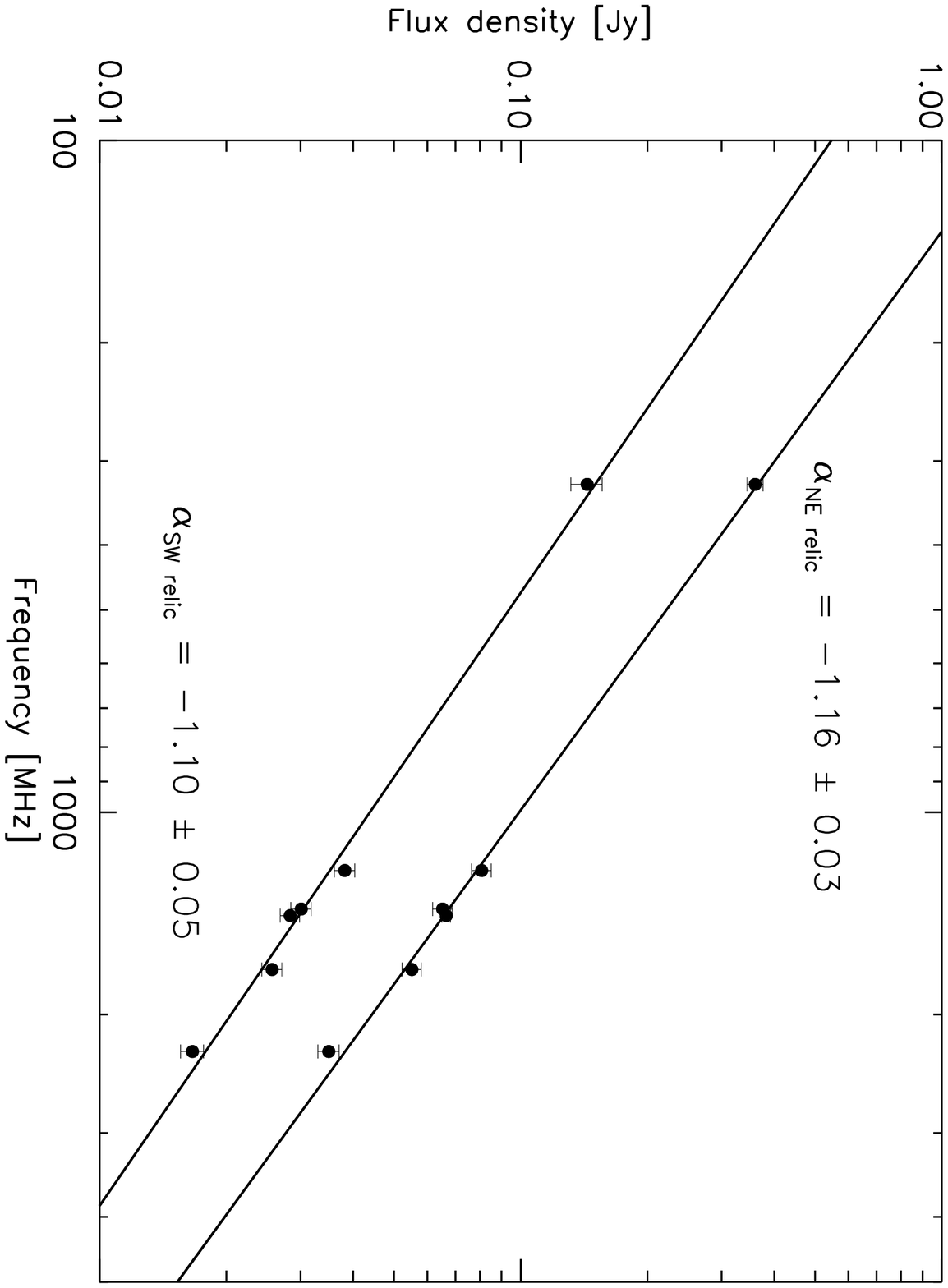}
\caption{ NVSS, WENSS and WSRT flux density measurements for the NE and SW relics. Power-law fits through these measurements are also shown} 
\label{fig:macsintsp}
\end{figure}

Integrated flux density measurements for the radio relics, from the WSRT, NVSS and WENSS images, are displayed in Fig.~\ref{fig:macsintsp}. A power-law fit through the measurements is also shown. This gives $\alpha=-1.16\pm0.03$ and $\alpha=-1.10\pm0.05$ for the NE and SW relic, respectively. The radio spectra appear to be straight over the observed frequency range.

\begin{table}
\begin{center}
\tabcolsep=0.1cm
\caption{Characteristics of the diffuse radio emission}
\begin{tabular}{llllll}
\hline
\hline
&  relic NE &  relic SW & radio halo  \\
\hline

$S_{1.714 \rm{\mbox{ }GHz}}$ (mJy)         & $55.1 \pm2.9$   & $25.7\pm1.4$ & $11.4 \pm 1.1$\\ 
$P_{1.4 \rm {\mbox{ }GHz}}$ ($10^{25}$ W Hz$^{-1}$)                  &  4.5  &  1.9 & 0.95$^{a}$\\
LLS$^{b}$ (kpc)	                 &  1340& 860 & $\sim1650$\\ 
 $\alpha$                    & $-1.16\pm0.03$ & $-1.10 \pm 0.05$ & \ldots\\
\hline
\hline
\end{tabular}
\label{tab:macs1752properties}
\end{center}
$^{a}$ assuming a spectral index of $-1.1$, $^{b}$ largest linear size
\end{table}

\section{Discussion}
\label{sec:discussion}
In the linear test particle regime for DSA, the injection radio spectral index ($\alpha_{\rm{inj}}$) is related to the Mach number ($\mathcal{M}$) of a shock \citep[e.g.,][]{1987PhR...154....1B}
\begin{equation}
\alpha_{\rm{inj}} =  \frac{1}{2} - \frac{\mathcal{M}^2 +1} {\mathcal{M}^2 -1} \mbox{ .}
\label{eq:inj-mach}
\end{equation}
If the properties (i.e., the Mach number) of the shock remain unchanged, the electron cooling time is shorter than the diffusion time, and the age of the relic is larger than the electron cooling time, the integrated radio spectrum will be a power-law, with a spectral index that is about 0.5 units steeper than $\alpha_{\rm{inj}}$ \citep{1962SvA.....6..317K}. The electron cooling time in clusters ($\sim 10^{6-7}$~yr at about 1~GHz), much shorter than the diffusion timescale \citep{1977ApJ...212....1J}. Also, the cooling time is likely shorter than the age of the relics in MACS~J1752.0+4440 since the shock waves probably formed at the moment of core passage of the two subclusters. The relics are located at about 800~kpc from the cluster center and it takes about $10^{8-9}$~yr for a shock wave to travel this distance, depending on the Mach number. This is more than an order of magnitude longer than the electron cooling time and therefore the Mach number has probably remained more or less constant over the last $10^{6-7}$~yr. We can thus assume $\alpha_{\rm{integrated}} = \alpha_{\rm{inj}} + 0.5$ and therefore $\alpha_{\rm{inj}}$ is expected to be around $-0.6$ to $-0.7$, similar to the $\alpha_{\rm{inj}}$ directly measured for the relics in CIZA~J2242.8+5301 \citep{2010Sci...330..347V} and 1RXS~J0603.3+4214 \citep{2011PhDT........14V}. 

These injection spectral indices correspond to Mach numbers of about 3.5 to 4.5. Quite high, but simulations show these can indeed be  created during cluster mergers \citep[e.g.,][]{2009A&A...504...33V,2010NewA...15..695V}.  Furthermore, Mach numbers derived from the radio spectral could be slightly higher than the hydrodynamical Mach number, if there are Mach number variations in 
the shock front \citep[since the radio luminosity strongly increases with Mach number, e.g.,][]{2007MNRAS.375...77H}.
Recent X-ray observations did reveal a $\mathcal{M} > 3$ shock, corresponding to the relic in CIZA~J2242.8+5301 \citep{2012arXiv1201.1502O,2011arXiv1112.3030A}. For MACS~J1752.0+4440, a next step will be to directly measure $\alpha_{\rm{inj}}$ at the front of the relic from a spectral index map. This requires additional observations which are currently obtained and the resulting spectral and polarization study will be presented in a forthcoming paper (Bonafede et al., MNRAS, submitted).

The presence of a radio halo in MACS~J1752.0+4440 provides support for the hypothesis that halos are related to cluster merger events. It also indicates that while the main shock waves from the merger are already located in the cluster outskirts, particle acceleration is still ongoing in the central parts of the cluster, as expected for the turbulent re-acceleration model.

The XMM image (Fig.~\ref{fig:vla}) suggests the cluster is undergoing a roughly equal mass binary merger event along in the NE-SW direction. The relative size of the two relics also gives an indication of the mass ratio of the merger event, while the brightness distribution along the  length of the relics is related to the impact parameter. Taking the results from the simulations presented in \cite{2011MNRAS.418..230V}, this suggest a mass ratio about $2:1$ and an impact parameter about 4 times the core-radius of the largest subcluster. These are very rough estimates though and the uncertainties are difficult to quantify. Moreover these simulations only hold under the assumptions of an isolated clean binary merger event (without any extra substructures). New simulations, in combination with X-ray and lensing analyses, are needed to produce a more detailed merger scenario. 

\section{Summary}
\label{sec:summary}
We reported the presence of a double radio relic system and radio halo in the X-ray luminous galaxy cluster MACS~J1752.0+4440 at $z=0.366$. This cluster is undergoing a binary cluster merger event, with the merger axis roughly located within the plane of the sky. For the radio halo we find a  luminosity of $P_{1.4 \rm { GHz}}=9.5\times10^{24}$~W~Hz$^{-1}$ and a total extent of 1.65~Mpc. The relics have sizes of about 0.9 and 1.3~Mpc. The integrated spectral indices of the relics are indicative of strong shock waves with $\mathcal{M}>3$, as is for the case for other recently discovered relics. This suggests that $\mathcal{M}>3$ shocks are not uncommon in the outskirts of merging clusters. 

\section*{Acknowledgments}
We would like to thank the anonymous referee for useful comments. 
AB, MB and MH acknowledge support by the research group FOR 1254 funded by the Deutsche Forschungsgemeinschaft. The Westerbork Synthesis Radio Telescope is operated by ASTRON (Netherlands Institute for Radio Astronomy) with support from the Netherlands Foundation for Scientific Research (NWO). The National Radio Astronomy Observatory is a facility of the National Science Foundation operated under cooperative agreement by Associated Universities, Inc. The NVAS image was produced as part of the NRAO VLA Archive Survey, (c) AUI/NRAO.
\bibliography{ref_filamentsmacs1752}

\begin{thebibliography}{63}
\expandafter\ifx\csname natexlab\endcsname\relax\def\natexlab#1{#1}\fi

\bibitem[{{Abazajian} {et~al}\mbox{.}(2009){Abazajian}, {Adelman-McCarthy},
  {Ag{\"u}eros}, {Allam}, {Allende Prieto}, {An}, {Anderson}, {Anderson},
  {Annis}, {Bahcall}, \& et~al.}]{2009ApJS..182..543A}
{Abazajian} K.~N. {et~al.}, 2009, \apjs, 182, 543

\bibitem[{{Akamatsu} \& {Kawahara}(2011)}]{2011arXiv1112.3030A}
{Akamatsu} H., {Kawahara} H., 2011, ArXiv e-prints

\bibitem[{{Baars} {et~al}\mbox{.}(1977){Baars}, {Genzel}, {Pauliny-Toth}, \&
  {Witzel}}]{1977A&A....61...99B}
{Baars} J.~W.~M., {Genzel} R., {Pauliny-Toth} I.~I.~K., {Witzel} A., 1977,
  \aap, 61, 99

\bibitem[{{Bagchi} {et~al}\mbox{.}(2011){Bagchi}, {Sirothia}, {Werner},
  {Pandge}, {Kantharia}, {Ishwara-Chandra}, {Gopal-Krishna}, {Paul}, \&
  {Joshi}}]{2011ApJ...736L...8B}
{Bagchi} J. {et~al.}, 2011, \apjl, 736, L8+

\bibitem[{{Blandford} \& {Eichler}(1987)}]{1987PhR...154....1B}
{Blandford} R., {Eichler} D., 1987, \physrep, 154, 1

\bibitem[{{Blasi} \& {Colafrancesco}(1999)}]{1999APh....12..169B}
{Blasi} P., {Colafrancesco} S., 1999, Astroparticle Physics, 12, 169

\bibitem[{{Bonafede} {et~al}\mbox{.}(2009){Bonafede}, {Feretti}, {Giovannini},
  {Govoni}, {Murgia}, {Taylor}, {Ebeling}, {Allen}, {Gentile}, \&
  {Pihlstr{\"o}m}}]{2009A&A...503..707B}
{Bonafede} A. {et~al.}, 2009, \aap, 503, 707

\bibitem[{{Brunetti} {et~al}\mbox{.}(2009){Brunetti}, {Cassano}, {Dolag}, \&
  {Setti}}]{2009A&A...507..661B}
{Brunetti} G., {Cassano} R., {Dolag} K., {Setti} G., 2009, \aap, 507, 661

\bibitem[{{Brunetti} {et~al}\mbox{.}(2008){Brunetti}, {Giacintucci}, {Cassano},
  {Lane}, {Dallacasa}, {Venturi}, {Kassim}, {Setti}, {Cotton}, \&
  {Markevitch}}]{2008Natur.455..944B}
{Brunetti} G. {et~al.}, 2008, \nat, 455, 944

\bibitem[{{Brunetti} {et~al}\mbox{.}(2001){Brunetti}, {Setti}, {Feretti}, \&
  {Giovannini}}]{2001MNRAS.320..365B}
{Brunetti} G., {Setti} G., {Feretti} L., {Giovannini} G., 2001, \mnras, 320,
  365

\bibitem[{{Brunetti} {et~al}\mbox{.}(2007){Brunetti}, {Venturi}, {Dallacasa},
  {Cassano}, {Dolag}, {Giacintucci}, \& {Setti}}]{2007ApJ...670L...5B}
{Brunetti} G., {Venturi} T., {Dallacasa} D., {Cassano} R., {Dolag} K.,
  {Giacintucci} S., {Setti} G., 2007, \apjl, 670, L5

\bibitem[{{Cassano} {et~al}\mbox{.}(2010{\natexlab{a}}){Cassano}, {Brunetti},
  {R{\"o}ttgering}, \& {Br{\"u}ggen}}]{2010A&A...509A..68C}
{Cassano} R., {Brunetti} G., {R{\"o}ttgering} H.~J.~A., {Br{\"u}ggen} M.,
  2010{\natexlab{a}}, \aap, 509, A68+

\bibitem[{{Cassano} {et~al}\mbox{.}(2006){Cassano}, {Brunetti}, \&
  {Setti}}]{2006MNRAS.369.1577C}
{Cassano} R., {Brunetti} G., {Setti} G., 2006, \mnras, 369, 1577

\bibitem[{{Cassano} {et~al}\mbox{.}(2007){Cassano}, {Brunetti}, {Setti},
  {Govoni}, \& {Dolag}}]{2007MNRAS.378.1565C}
{Cassano} R., {Brunetti} G., {Setti} G., {Govoni} F., {Dolag} K., 2007, \mnras,
  378, 1565

\bibitem[{{Cassano} {et~al}\mbox{.}(2010{\natexlab{b}}){Cassano}, {Ettori},
  {Giacintucci}, {Brunetti}, {Markevitch}, {Venturi}, \&
  {Gitti}}]{2010ApJ...721L..82C}
{Cassano} R., {Ettori} S., {Giacintucci} S., {Brunetti} G., {Markevitch} M.,
  {Venturi} T., {Gitti} M., 2010{\natexlab{b}}, \apjl, 721, L82

\bibitem[{{Condon} {et~al}\mbox{.}(1998){Condon}, {Cotton}, {Greisen}, {Yin},
  {Perley}, {Taylor}, \& {Broderick}}]{1998AJ....115.1693C}
{Condon} J.~J., {Cotton} W.~D., {Greisen} E.~W., {Yin} Q.~F., {Perley} R.~A.,
  {Taylor} G.~B., {Broderick} J.~J., 1998, \aj, 115, 1693

\bibitem[{{Dennison}(1980)}]{1980ApJ...239L..93D}
{Dennison} B., 1980, \apjl, 239, L93

\bibitem[{{Dolag} \& {En{\ss}lin}(2000)}]{2000A&A...362..151D}
{Dolag} K., {En{\ss}lin} T.~A., 2000, \aap, 362, 151

\bibitem[{{Drury}(1983)}]{1983RPPh...46..973D}
{Drury} L.~O., 1983, Reports on Progress in Physics, 46, 973

\bibitem[{{Ebeling} {et~al}\mbox{.}(2004){Ebeling}, {Barrett}, \&
  {Donovan}}]{2004ApJ...609L..49E}
{Ebeling} H., {Barrett} E., {Donovan} D., 2004, \apjl, 609, L49

\bibitem[{{Ebeling} {et~al}\mbox{.}(2001){Ebeling}, {Edge}, \&
  {Henry}}]{2001ApJ...553..668E}
{Ebeling} H., {Edge} A.~C., {Henry} J.~P., 2001, \apj, 553, 668

\bibitem[{{Ebeling et al.}(2012)}]{ebeling}
{Ebeling et al.}, 2012, in preparation

\bibitem[{{Edge} {et~al}\mbox{.}(2003){Edge}, {Ebeling}, {Bremer},
  {R{\"o}ttgering}, {van Haarlem}, {Rengelink}, \&
  {Courtney}}]{2003MNRAS.339..913E}
{Edge} A.~C., {Ebeling} H., {Bremer} M., {R{\"o}ttgering} H., {van Haarlem}
  M.~P., {Rengelink} R., {Courtney} N.~J.~D., 2003, \mnras, 339, 913

\bibitem[{{En{\ss}lin} {et~al}\mbox{.}(2011){En{\ss}lin}, {Pfrommer},
  {Miniati}, \& {Subramanian}}]{2011A&A...527A..99E}
{En{\ss}lin} T., {Pfrommer} C., {Miniati} F., {Subramanian} K., 2011, \aap,
  527, A99+

\bibitem[{{En{\ss}lin} \& {R{\"o}ttgering}(2002)}]{2002A&A...396...83E}
{En{\ss}lin} T.~A., {R{\"o}ttgering} H., 2002, \aap, 396, 83

\bibitem[{{Feretti} {et~al}\mbox{.}(2012){Feretti}, {Giovannini}, {Govoni}, \&
  {Murgia}}]{2012arXiv1205.1919F}
{Feretti} L., {Giovannini} G., {Govoni} F., {Murgia} M., 2012, ArXiv e-prints

\bibitem[{{Giacintucci} {et~al}\mbox{.}(2008){Giacintucci}, {Venturi},
  {Macario}, {Dallacasa}, {Brunetti}, {Markevitch}, {Cassano}, {Bardelli}, \&
  {Athreya}}]{2008A&A...486..347G}
{Giacintucci} S. {et~al.}, 2008, \aap, 486, 347

\bibitem[{{Giovannini} {et~al}\mbox{.}(2009){Giovannini}, {Bonafede},
  {Feretti}, {Govoni}, {Murgia}, {Ferrari}, \& {Monti}}]{2009A&A...507.1257G}
{Giovannini} G., {Bonafede} A., {Feretti} L., {Govoni} F., {Murgia} M.,
  {Ferrari} F., {Monti} G., 2009, \aap, 507, 1257

\bibitem[{{Hoeft} \& {Br{\"u}ggen}(2007)}]{2007MNRAS.375...77H}
{Hoeft} M., {Br{\"u}ggen} M., 2007, \mnras, 375, 77

\bibitem[{{Jaffe}(1977)}]{1977ApJ...212....1J}
{Jaffe} W.~J., 1977, \apj, 212, 1

\bibitem[{{Jeltema} \& {Profumo}(2011)}]{2011ApJ...728...53J}
{Jeltema} T.~E., {Profumo} S., 2011, \apj, 728, 53

\bibitem[{{Jones} \& {Ellison}(1991)}]{1991SSRv...58..259J}
{Jones} F.~C., {Ellison} D.~C., 1991, Space Science Reviews, 58, 259

\bibitem[{{Kang} \& {Ryu}(2011)}]{2011ApJ...734...18K}
{Kang} H., {Ryu} D., 2011, \apj, 734, 18

\bibitem[{{Kardashev}(1962)}]{1962SvA.....6..317K}
{Kardashev} N.~S., 1962, Soviet Astronomy, 6, 317

\bibitem[{{Kartaltepe} {et~al}\mbox{.}(2008){Kartaltepe}, {Ebeling}, {Ma}, \&
  {Donovan}}]{2008MNRAS.389.1240K}
{Kartaltepe} J.~S., {Ebeling} H., {Ma} C.~J., {Donovan} D., 2008, \mnras, 389,
  1240

\bibitem[{{LaRoque} {et~al}\mbox{.}(2003){LaRoque}, {Joy}, {Carlstrom},
  {Ebeling}, {Bonamente}, {Dawson}, {Edge}, {Holzapfel}, {Miller}, {Nagai},
  {Patel}, \& {Reese}}]{2003ApJ...583..559L}
{LaRoque} S.~J. {et~al.}, 2003, \apj, 583, 559

\bibitem[{{Liang} {et~al}\mbox{.}(2000){Liang}, {Hunstead}, {Birkinshaw}, \&
  {Andreani}}]{2000ApJ...544..686L}
{Liang} H., {Hunstead} R.~W., {Birkinshaw} M., {Andreani} P., 2000, \apj, 544,
  686

\bibitem[{{Limousin} {et~al}\mbox{.}(2011){Limousin}, {Ebeling}, {Richard},
  {Swinbank}, {Smith}, {Rodionov}, {Ma}, {Smail}, {Edge}, {Jauzac}, {Jullo}, \&
  {Kneib}}]{2011arXiv1109.3301L}
{Limousin} M. {et~al.}, 2011, ArXiv e-prints

\bibitem[{{Ma} {et~al}\mbox{.}(2009){Ma}, {Ebeling}, \&
  {Barrett}}]{2009ApJ...693L..56M}
{Ma} C., {Ebeling} H., {Barrett} E., 2009, \apjl, 693, L56

\bibitem[{{Ma} \& {Ebeling}(2011)}]{2011MNRAS.410.2593M}
{Ma} C.-J., {Ebeling} H., 2011, \mnras, 410, 2593

\bibitem[{{Ma} {et~al}\mbox{.}(2008){Ma}, {Ebeling}, {Donovan}, \&
  {Barrett}}]{2008ApJ...684..160M}
{Ma} C.-J., {Ebeling} H., {Donovan} D., {Barrett} E., 2008, \apj, 684, 160

\bibitem[{{Macario} {et~al}\mbox{.}(2011){Macario}, {Markevitch},
  {Giacintucci}, {Brunetti}, {Venturi}, \& {Murray}}]{2011ApJ...728...82M}
{Macario} G., {Markevitch} M., {Giacintucci} S., {Brunetti} G., {Venturi} T.,
  {Murray} S.~S., 2011, \apj, 728, 82

\bibitem[{{Mann} \& {Ebeling}(2012)}]{2012MNRAS.420.2120M}
{Mann} A.~W., {Ebeling} H., 2012, \mnras, 420, 2120

\bibitem[{{Markevitch} {et~al}\mbox{.}(2005){Markevitch}, {Govoni}, {Brunetti},
  \& {Jerius}}]{2005ApJ...627..733M}
{Markevitch} M., {Govoni} F., {Brunetti} G., {Jerius} D., 2005, \apj, 627, 733

\bibitem[{{Murgia} {et~al}\mbox{.}(2009){Murgia}, {Govoni}, {Markevitch},
  {Feretti}, {Giovannini}, {Taylor}, \& {Carretti}}]{2009A&A...499..679M}
{Murgia} M., {Govoni} F., {Markevitch} M., {Feretti} L., {Giovannini} G.,
  {Taylor} G.~B., {Carretti} E., 2009, \aap, 499, 679

\bibitem[{{Nuza} {et~al}\mbox{.}(2012){Nuza}, {Hoeft}, {van Weeren},
  {Gottl{\"o}ber}, \& {Yepes}}]{2012MNRAS.420.2006N}
{Nuza} S.~E., {Hoeft} M., {van Weeren} R.~J., {Gottl{\"o}ber} S., {Yepes} G.,
  2012, \mnras, 420, 2006

\bibitem[{{Offringa} {et~al}\mbox{.}(2010){Offringa}, {de Bruyn}, {Biehl},
  {Zaroubi}, {Bernardi}, \& {Pandey}}]{2010MNRAS.405..155O}
{Offringa} A.~R., {de Bruyn} A.~G., {Biehl} M., {Zaroubi} S., {Bernardi} G.,
  {Pandey} V.~N., 2010, \mnras, 405, 155

\bibitem[{{Ogrean} {et~al}\mbox{.}(2012){Ogrean}, {Br{\"u}ggen},
  {R{\"o}ttgering}, {Simionescu}, {Croston}, {van Weeren}, \&
  {Hoeft}}]{2012arXiv1201.1502O}
{Ogrean} G., {Br{\"u}ggen} M., {R{\"o}ttgering} H., {Simionescu} A., {Croston}
  J., {van Weeren} R., {Hoeft} M., 2012, ArXiv e-prints

\bibitem[{{Perley} \& {Taylor}(1999)}]{perleyandtaylor}
{Perley} R.~T., {Taylor} G.~B., 1999, {VLA Calibrator Manual}. Tech. rep., NRAO

\bibitem[{{Petrosian}(2001)}]{2001ApJ...557..560P}
{Petrosian} V., 2001, \apj, 557, 560

\bibitem[{{Planck Collaboration} {et~al}\mbox{.}(2011){Planck Collaboration},
  {Ade}, {Aghanim}, {Arnaud}, {Ashdown}, {Aumont}, {Baccigalupi}, {Balbi},
  {Banday}, {Barreiro}, \& et~al.}]{2011A&A...536A...8P}
{Planck Collaboration} {et~al.}, 2011, \aap, 536, A8

\bibitem[{{Rengelink} {et~al}\mbox{.}(1997){Rengelink}, {Tang}, {de Bruyn},
  {Miley}, {Bremer}, {R\"ottgering}, \& {Bremer}}]{1997A&AS..124..259R}
{Rengelink} R.~B., {Tang} Y., {de Bruyn} A.~G., {Miley} G.~K., {Bremer} M.~N.,
  {R\"ottgering} H.~J.~A., {Bremer} M.~A.~R., 1997, \aaps, 124, 259

\bibitem[{{Roettiger} {et~al}\mbox{.}(1999){Roettiger}, {Burns}, \&
  {Stone}}]{1999ApJ...518..603R}
{Roettiger} K., {Burns} J.~O., {Stone} J.~M., 1999, \apj, 518, 603

\bibitem[{{van Weeren}(2011)}]{2011PhDT........14V}
{van Weeren} R.~J., 2011, PhD thesis, Leiden Observatory, The Netherlands

\bibitem[{{van Weeren} {et~al}\mbox{.}(2011){van Weeren}, {Br{\"u}ggen},
  {R{\"o}ttgering}, \& {Hoeft}}]{2011MNRAS.418..230V}
{van Weeren} R.~J., {Br{\"u}ggen} M., {R{\"o}ttgering} H.~J.~A., {Hoeft} M.,
  2011, \mnras, 418, 230

\bibitem[{{van Weeren} {et~al}\mbox{.}(2009){van Weeren}, {R{\"o}ttgering},
  {Br{\"u}ggen}, \& {Cohen}}]{2009A&A...505..991V}
{van Weeren} R.~J., {R{\"o}ttgering} H.~J.~A., {Br{\"u}ggen} M., {Cohen} A.,
  2009, \aap, 505, 991

\bibitem[{{van Weeren} {et~al}\mbox{.}(2010){van Weeren}, {R{\"o}ttgering},
  {Br{\"u}ggen}, \& {Hoeft}}]{2010Sci...330..347V}
{van Weeren} R.~J., {R{\"o}ttgering} H.~J.~A., {Br{\"u}ggen} M., {Hoeft} M.,
  2010, Science, 330, 347

\bibitem[{{van Weeren} {et~al}\mbox{.}(2012){van Weeren}, {Rottgering},
  {Rafferty}, {Pizzo}, {Bonafede}, {Bruggen}, {Brunetti}, {Ferrari}, {Orru},
  {Heald}, {McKean}, {Tasse}, {de Gasperin}, {Birzan}, {van Zwieten}, {van der
  Tol}, {Shulevski}, {Jackson}, {Offringa}, {Conway}, {Intema}, {Clarke}, {van
  Bemmel}, {Miley}, {White}, {Hoeft}, {Cassano}, {Macario}, {Morganti}, {Wise},
  {Horellou}, {Valentijn}, {Wucknitz}, {Kuijken}, {Ensslin}, {Anderson},
  {Asgekar}, {Avruch}, {Beck}, {Bell}, {Bell}, {Bentum}, {Bernardi}, {Best},
  {Boonstra}, {Brentjens}, {van de Brink}, {Broderick}, {Brouw}, {Butcher},
  {van Cappellen}, {Ciardi}, {Eisloffel}, {Falcke}, {Fender}, {Garrett},
  {Gerbers}, {Gunst}, {Hassall}, {Hessels}, {Koopmans}, {Kuper}, {van Leeuwen},
  {Maat}, {Millenaar}, {Munk}, {Nijboer}, {Noordam}, {Pandey},
  {Pandey-Pommier}, {Polatidis}, {Reich}, {Scaife}, {Schoenmakers}, {Sluman},
  {Stappers}, {Steinmetz}, {Swinbank}, {Tagger}, {Tang}, {Vermeulen}, \& {de
  Vos}}]{2012arXiv1205.4730V}
{van Weeren} R.~J. {et~al.}, 2012, ArXiv e-prints

\bibitem[{{Vazza} {et~al}\mbox{.}(2012){Vazza}, {Br{\"u}ggen}, {van Weeren},
  {Bonafede}, {Dolag}, \& {Brunetti}}]{2012MNRAS.421.1868V}
{Vazza} F., {Br{\"u}ggen} M., {van Weeren} R., {Bonafede} A., {Dolag} K.,
  {Brunetti} G., 2012, \mnras, 421, 1868

\bibitem[{{Vazza} {et~al}\mbox{.}(2010){Vazza}, {Brunetti}, {Gheller}, \&
  {Brunino}}]{2010NewA...15..695V}
{Vazza} F., {Brunetti} G., {Gheller} C., {Brunino} R., 2010, New Astronomy, 15,
  695

\bibitem[{{Vazza} {et~al}\mbox{.}(2009){Vazza}, {Brunetti}, {Kritsuk},
  {Wagner}, {Gheller}, \& {Norman}}]{2009A&A...504...33V}
{Vazza} F., {Brunetti} G., {Kritsuk} A., {Wagner} R., {Gheller} C., {Norman}
  M., 2009, \aap, 504, 33

\bibitem[{{Voges} {et~al}\mbox{.}(1999){Voges}, {Aschenbach}, {Boller},
  {Br{\"a}uninger}, {Briel}, {Burkert}, {Dennerl}, \& {et
  al.}}]{1999A&A...349..389V}
{Voges} W., {Aschenbach} B., {Boller} T., {Br{\"a}uninger} H., {Briel} U.,
  {Burkert} W., {Dennerl} K., {et al.}, 1999, \aap, 349, 389

\bibitem[{{Zitrin} {et~al}\mbox{.}(2009){Zitrin}, {Broadhurst}, {Rephaeli}, \&
  {Sadeh}}]{2009ApJ...707L.102Z}
{Zitrin} A., {Broadhurst} T., {Rephaeli} Y., {Sadeh} S., 2009, \apjl, 707, L102

\end{thebibliography}

\label{lastpage}

\end{document}